\documentclass[aps,prl,twocolumn,showpacs,superscriptaddress]{revtex4-1}
\usepackage{dcolumn}
\usepackage{graphicx}
\usepackage{array}
\usepackage[utf8x]{inputenc}
\usepackage{xcolor}
\usepackage{float}
\usepackage{natbib}
\usepackage{bm}
\usepackage[T1]{fontenc}
\usepackage{mathptmx}
\usepackage{physics}
\usepackage{amsmath}
\usepackage{nicefrac}
\usepackage{todonotes}
\usepackage{array}
\usepackage{makecell}
\usepackage{csquotes}
\usepackage{hyperref}
\usepackage{upgreek}
\usepackage{mathtools, nccmath}
\usepackage{lipsum}
\usepackage[normalem]{ulem}
\usepackage{hyperref}
\usepackage[mathlines]{lineno}

\def \vns{V$_{1/3}$NbS$_2$}
\def \cm{cm$^{-1}$}

\begin{document}
\preprint{APS/123-QED}
\title{Raman spectroscopic evidence for linearly dispersed nodes and magnetic ordering in the topological semimetal  V$_{1/3}$NbS$_2$}
\author{Shreenanda Ghosh}
\email{shreenanda@jhu.edu}
\affiliation{Institute for Quantum Matter and William H. Miller III Department of Physics and Astronomy, Johns Hopkins University, Baltimore, Maryland 21218, USA}
\author{Chris Lygouras}
\affiliation{Institute for Quantum Matter and William H. Miller III Department of Physics and Astronomy, Johns Hopkins University, Baltimore, Maryland 21218, USA}
\author{Zili Feng}
\affiliation{Department of Physics, University of Tokyo, Bunkyo-ku, Tokyo 113-0033, Japan}
\affiliation{Institute for Solid State Physics, University of Tokyo, Kashiwa, Chiba 277-8581, Japan}
\affiliation{Trans-scale Quantum Science Institute, University of Tokyo, Bunkyo-ku, Tokyo 113-0033, Japan}
\author{Mingxuan Fu}
\affiliation{Department of Physics, University of Tokyo, Bunkyo-ku, Tokyo 113-0033, Japan}
\affiliation{Institute for Solid State Physics, University of Tokyo, Kashiwa, Chiba 277-8581, Japan}
\author{Satoru Nakatsuji}
\affiliation{Institute for Quantum Matter and William H. Miller III Department of Physics and Astronomy, Johns Hopkins University, Baltimore, Maryland 21218, USA}
\affiliation{Department of Physics, University of Tokyo, Bunkyo-ku, Tokyo 113-0033, Japan}
\affiliation{Institute for Solid State Physics, University of Tokyo, Kashiwa, Chiba 277-8581, Japan}
\affiliation{Trans-scale Quantum Science Institute, University of Tokyo, Bunkyo-ku, Tokyo 113-0033, Japan}
\author{Natalia Drichko}
\email{drichko@jhu.edu}
\affiliation{Institute for Quantum Matter and William H. Miller III Department of Physics and Astronomy, Johns Hopkins University, Baltimore, Maryland 21218, USA}

\date{\today}

\begin{abstract}
Weyl semimetals are characterized by an electronic structure with linearly dispersed nodes and distinguished chirality, protected by broken inversion or time reversal symmetry.  The intercalated transition metal dichalcogenide V$_{1/3}$NbS$_2$ is proposed as a Weyl semimetal. In this study, we report polarization-resolved magnetic and electronic Raman scattering of this material, probing both the magnetic order and the electronic structure. The electronic scattering reveals a linear with frequency continuum of excitations, as the signature of electronic transitions within the proposed Weyl nodes in a two-dimensional electronic structure. Additionally, two-magnon excitations of V moments are observed near 15 meV in the magnetically ordered phase below 50 K. These excitations are well reproduced by calculations based on the Fleury-Loudon theory using spin wave exchange parameters derived from the neutron scattering data of this material and confirm the antiferromagnetic character of the order. These magnetic and electronic scattering, observed in the same spectra, provide independent spectroscopic evidence for a collinear antiferromagnetic Weyl semimetal state in V$_{1/3}$NbS$_2$.
\end{abstract}

\maketitle


Despite early recognition in the last century, solid state physics \cite{Herring1937}, topological semimetals, characterized by the occurrence of band touching points or nodes at the Fermi energy ($E_{\rm F}$), have attracted significant interest only during the past decade. Among these, Weyl semimetals (WSM) have fascinating properties. Numerous theoretical and experimental investigations have been performed \cite{Burkov2016,Binghai2017, Armitage2018,nakatsuji2022topological}, to study their unusual transport properties, including the anomalous Hall effect (AHE) \cite{nakatsuji2015large,Liu2016, Liu2018, TChen2021mn3x}, the unconventional magnetoresistance \cite{Shekhar2015, Ian2018}, and the chiral anomaly \cite{Aji2012,kuroda2017evidence, Liu2013,Ong2021}. Optical phonons in WSM are also a focus due to their interaction with the non-trivial electronic band structure 
\cite{Song2016,Rinkel2019}.

In the presence of time reversal symmetry breaking, in particular, the topologically non-trivial states play a pivotal role in generating the anomalous transverse response and novel functionalities \cite{nakatsuji2022topological,vsmejkal2022anomalous}. Such novel properties in topological magnets have been highlighted by the observation of large anomalous transverse responses beyond their scaling law with magnetization. Recent examples are the observation of the large AHE in the non-coplanar frustrated magnet Pr$_2$Ir$_2$O$_7$ and the non-collinear antiferromagnet Mn$_3$Sn \cite{machida2010time,nakatsuji2015large,ikhlas2017large,TChen2021mn3x}. The chiral antiferromagnet Mn$_3$Sn hosts Weyl points near the $E_{\rm F}$, and the strongly enhanced Berry curvature generates gigantic transverse responses, despite vanishingly small magnetization \cite{kuroda2017evidence,TChen2021mn3x}. 

Recently, symmetry analyses have revealed that several collinear antiferromagnets, called altermagnets, may exhibit non-zero anomalous transverse responses despite the nearly absent magnetization \cite{vsmejkal2022anomalous,Altermagnet1,Sinova2022}. However, given the magnetization scaling law, the size of these responses is found to be relatively small compared to the actual value of the magnetization in the altermagnets reported thus far \cite{MnTeAHE2023,Mn5SiAHE2024,Takagi2025}. In this sense, the discovery of a large AHE in the collinear antiferromagnetic (AFM) state of V$_{1/3}$NbS$_2$ is striking and suggests a topologically nontrivial electronic structure behind the response beyond the scaling law \cite{Youzhe2025}. In fact, experimental and theoretical studies have indicated that V$_{1/3}$NbS$_2$ may host a Weyl semimetallic state \cite{Inoshita2019,Youzhe2025} and thus harbor large Berry curvature near the $E_{\rm F}$.

\par 
High resolution is necessary for the spectroscopic characterization of topologically nontrivial electron bands in WSM, especially for confirming the presence of Weyl nodes near the $E_{\rm F}$. The case of pyrochlore iridates, a family of topological magnets that are expected to host Weyl and Luttinger semimetal states, has been particularly illuminating in this regard \cite{machida2010time,kondo2015}. While evidence for Weyl fermions could not be obtained with angle-resolved photoemission spectroscopy due to the low-energy resolution \cite{Nakayama2016}, by calculating the Raman response using perturbation theory and comparing it with electronic contributions to Raman scattering, evidence of Weyl and Luttinger quasi-particles has been identified across various temperature ranges \cite{Niko2024}. A similar way of probing the dispersion through optical excitations can be provided by infrared and terahertz spectroscopy \cite{Sushkov2015}, but it is limited in probing low frequencies by sample sizes and is not sensitive to the excitations of the magnetic system. 

\begin{figure}
\includegraphics[width=\columnwidth]{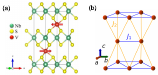}
\caption{\textbf{(a)} Crystal structure of \vns, which crystallizes in the hexagonal, chiral, noncentrosymmetric P6$_3$22 space group, viewed along the $\vec{a*}$ direction. V atoms are shown with arrows depicting the antiferromagnetically aligned spins in the neighboring planes. This figure was made using VESTA\cite{Momma2011}.
\textbf{(b)} Simplified structure showing the  sublattice of  V atoms, and magnetic exchange interactions: in-plane  nearest-neighbor $J_1 < $0, intralayer  $J_2 >$ 0.} \label{fig:struc}
\end{figure}

Similar to other X$_{1/3}$NbS$_2$(X= Cr, Mn, Fe, Co, Ni, V) magnetically intercalated transition metal dichalcogenide (TMDC) compounds of the same family, V$_{1/3}$NbS$_2$ crystallizes without inversion or mirror symmetry. The crystal structure is shown in Fig. \ref{fig:struc}(a),  where  S-Nb-S layers are stacked along the ${c}$-axis, with the V atoms lying in between the S-Nb-S layers.  A unit cell consists of two intercalant V layers.  
Hybrid-functional-based ab initio calculations proposed  V$_{1/3}$NbS$_2$ to be a ferromagnetic (FM) WSM \cite{Inoshita2019}, with only two bands present at the $E_{\rm F}$ and crossing each other to form Weyl points, in the absence of any spin-orbit interaction. Another recent first-principles study using spin-resolved Density functional theory (DFT) calculations also predicted V$_{1/3}$NbS$_2$ to host a ferromagnetically ordered state \cite{Hawkhead2023}. On the other hand, more recent first-principles calculations based on spin DFT predicted an AFM structure \cite{Hatanaka2023}. This disparity between the expected magnetic structures from band structure calculations calls for experiments that can resolve it.

Experimental efforts to determine the magnetic structure in  V$_{1/3}$NbS$_2$  have also contributed to the ongoing debate. It was originally reported to be a paramagnet \cite{HULLIGER1970}, followed by magnetic susceptibility measurements finding it to order ferromagnetically at temperatures below 50 K \cite{Parkin1980}. However, according to the recent neutron scattering and diffraction studies, the magnetic moments of the intercalant V form a collinear A-type order, where the triangular superlattice of V atoms is ferromagnetically ordered within the layer and antiferromagnetically ordered between the layers \cite{Lu2020, Hall2021, Youzhe2025}. 
This magnetic order makes V$_{1/3}$NbS$_2$ a candidate material for altermagnetism \cite{vsmejkal2022anomalous,Sinova2022}. The details of the magnetic order directly influence the band structure and band crossings; therefore, studying the magnetic structure and its interplay with electronic properties through spectroscopic probes will be crucial for understanding the ground state in V$_{1/3}$NbS$_2$.

Raman scattering spectroscopy can probe electronic and magnetic excitations in a single experiment, allowing us to identify Weyl nodes' dispersion and the onset of magnetic order. In this Letter, using polarization-resolved Raman spectroscopy, we observe a continuum of electronic Raman scattering with linear frequency dependence in V$_{1/3}$NbS$_2$, consistent with electronic transitions occurring in a Weyl node, hosted by a two-dimensional (2D) electronic structure. In the magnetically ordered phase below the N\'eel temperature $T_{\rm N}$ $\approx$ 50~K, we also detect a broad magnetic excitation near 15 meV, originating from the two-magnon (2M) scattering of vanadium $S = 1$ ordered moments. The 2M excitation is well-reproduced by calculations using  Fleury-Loudon (FL) scattering \cite{Fleury1968} and linear spin wave theory. This provides an independent spectroscopic confirmation of the AFM order in this material, in agreement with the order proposed by neutron diffraction measurements \cite{Lu2020, Youzhe2025}. These results on electronic and magnetic Raman scattering observed for \vns\ provide evidence of the electronic structure and magnetism necessary to realize a Weyl semimetal state in this material. 

\begin{table}[ht!]
\centering
\small 
\setlength{\tabcolsep}{6pt} 
\renewcommand{\arraystretch}{1.2} 
\begin{tabular}{>{\centering\arraybackslash}m{1 cm} >{\centering\arraybackslash}m{1 cm} >{\centering\arraybackslash}m{1.5 cm} >{\centering\arraybackslash}m{1 cm}}
\hline\hline
\thead{Frequency \\ (cm$^{-1}$)} & \thead{Symmetry} & \thead{linewidth in XX \\ (cm$^{-1}$)} & \thead{linewidth in ZZ \\ (cm$^{-1}$)} \\
\hline 202 & $A_1$ & 5 & 4 \\ 340 & $A_1$ & 6 & 5 \\ 406 & $A_1$ & 7 & 5 \\
\hline\hline
\end{tabular}
\caption{Comparison of frequencies and linewidth for in-plane and out-of-plane phonons, as measured at room temperature.} \label{table:phonons} \end{table}

\begin{figure}
\includegraphics[width=0.5\textwidth]{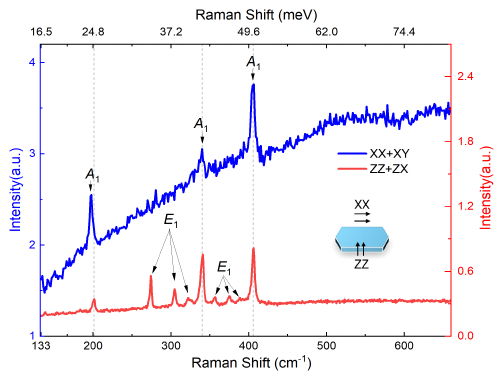}
\caption{Room temperature Raman spectra of \vns, measured along in-plane and out-of-plane configuration using micro-Raman. The schematic shows the sample configuration and measured polarizations. The symmetries of the observed phonons are marked in the plot. }\label{fig:outofplane}
\end{figure}

The room temperature measurements of Raman scattering spectra of a single crystal of V$_{1/3}$NbS$_2$ in the in-$(ab)$-plane ${(xx)+(xy)}$  and out-of-$(ab)$-plane  $(zz)+(zx)$ scattering geometries allow us to probe the anisotropy of the electronic Raman scattering,
which would be expected for a TMDC material, see Fig.~\ref{fig:outofplane}. Due to the small size of the single crystals, with a thickness of about 50~$\mu$m, these measurements in the out-of-plane ($(zz)+(zx)$) geometry are possible only with a micro-Raman setup (see \hyperref[sec:SI]{SI} for details of the setup) in the frequency range down to 133~\cm only.

\begin{figure}
\includegraphics[width = 0.5\textwidth]{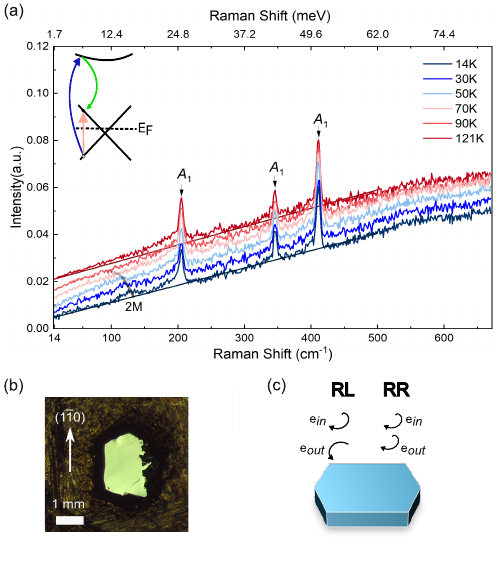}
\caption{\textbf{(a)} Temperature dependent Raman spectra of V$_{1/3}$NbS$_2$ measured in RR scattering channel excited by 514.5 nm. For clarity, each spectrum except the 14 K has been shifted along the y-axis. The translucent line is a guide to the eye to follow the temperature dependence of the 2M excitation. The inset is a schematic diagram showing the Raman scattering process within the Weyl node. \textbf{(b)} A photograph of the sample showing the top view. \textbf{(c)} Schematic of the sample and the circular polarization configuration used for the measurements at cryogenic temperatures is shown. $\vec{e}_\mathrm{in}$ and $\vec{e}_\mathrm{out}$ are the polarizations of the incident and the scattered light, respectively. R(L) denotes right(left) circularly polarized light.}\label{fig:phonons}
\end{figure}

The in-$(ab)$-plane  ${(xx)+(xy)}$ spectra demonstrate a strong continuum, absent in the $(zz)$ spectra, which we attribute to the electronic scattering discussed later. The narrow excitations observed in both spectra are phonons, summarized in Table \ref{table:phonons}. The phonons at 202, 340, and 406 cm$^{-1}$ observed in both ${(xx)+(xy)}$ and $(zz)$ channels with varying intensity belong to $A_1$ symmetry,  $E_2$ phonons expected in the $xy$ and RL in-plane scattering channels are too weak to be detected. In the out-of-$(ab)$-plane scattering spectra $(zz)+(zx)$,  weak features of $E_1$ phonons are observed, summarized in Table \ref{table:E1phonons}. A comparison with isostructural 2H-MoS$_2$ allows us to assign phonons observed in this spectral range to the sulfur displacements \cite{Kim2020}.

While the frequencies of $A_1$ phonons are similar in both channels, the lineshape differs between in-plane and out-of-plane polarized scattering: all three $A_1$ phonons are slightly broader in ${(xx)+(xy)}$ (see Table~\ref{table:phonons}), with the phonon at 406 cm$^{-1}$ demonstrating the asymmetric Fano shape. The additional broadening of the phonons and the Fano shape suggest coupling with electronic excitations \cite{Fano1961,He2024}. The frequency of the lowest energy phonon at 202 cm$^{-1}$ for out-of-$(ab)$-plane polarized scattering is lowered by 4 wave numbers in the in-$(ab)$-plane spectra. Such lowering of phonon frequency is also characteristic of electron-phonon interactions \cite {Cong2020}.

The temperature dependence of the Raman scattering spectra of \vns\ in the RR scattering channel in-$ab$-plane of the crystal was measured in 13-670 cm$^{-1}$ spectral range down to 14~K, see Fig. \ref{fig:phonons}(a). The phonon modes exhibit slight hardening on lowering the temperature without any other changes, supporting other reports on the absence of a structural phase transition at $T_{\rm N}$ \cite{Parkin1980, Lu2020, Hall2021}. The details of the temperature dependence of the phonon parameters are presented in Fig. \ref{fig:parameters} of sec \hyperref[sec:SI]{SI}.

\begin{figure}
\includegraphics[width = 0.5\textwidth]{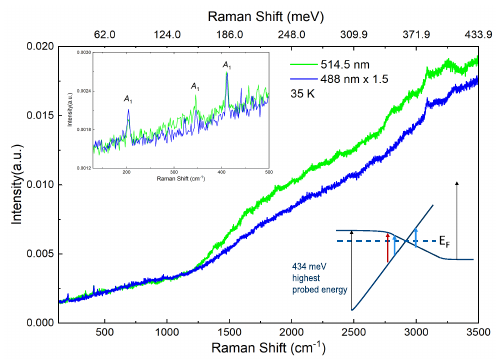}
\caption{Raman susceptibility over extended range using 488 nm and 514.5 nm excitation lines in RR scattering channel, at 35 K. The laser power varied between 10 (for 514.5 nm) - 15 mW (for 488 nm). The inset at the top left shows the $A_1$ phonons probed by different excitation lines. The inset at the bottom right shows a schematic view of the probed band structure. The blue arrows represent probing the Weyl nodes at low energies, and the change in the dispersion is pointed out by the red arrow.
}\label{fig:Extrange}
\end{figure}

The Raman spectra in the in-$(ab)$-plane scattering channels at all temperatures show a continuum of excitations increasing linearly with frequency up to 500~\cm (see Fig. \ref{fig:phonons}) and some deviation from linear behavior above this frequency. The origin of this continuum as Raman scattering is confirmed by a comparison of measurements with 514.5 nm, and 488 nm excitation laser lines (see Fig. \ref{fig:Extrange}), which demonstrates that the deviation between these two spectra occurs above 1300~\cm, most probably due to the additional weak photoluminescence excited with 514.5 nm line. 
The linear dependence of the Raman scattering continuum on frequency does not show any overall changes as the temperature decreases from 121 K to 14 K.  

\begin{figure}
\includegraphics[width = 0.5\textwidth]{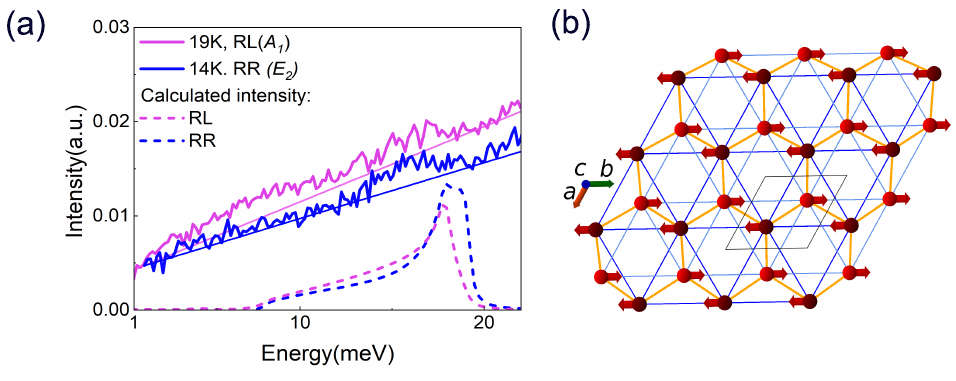}
\caption{\textbf{(a)} Raman scattering spectra below $T_{\rm N}$ in RL and RR polarization, versus calculated 2M intensity (dashed lines). The overall energy scale of the calculated intensity of 2M excitations matches the experimental observations. \textbf{(b)} View of the crystal structure from the c-axis, showing the in-plane triangular lattice ferromagnetic interactions between V atoms in blue and out-of-plane AFM interactions in orange. A  non-zero projection of the inter-layer AFM exchange onto the $(ab)$ plane allows a non-zero 2M  Raman scattering intensity following the Fleury-Loudon Raman process.}\label{fig:magnon}

\label{fig:magnon_new}
\end{figure}

{\it Two-magnon Raman scattering from V magnetic order} Below $T_{\rm N}$ = 50~K, we detect an additional excitation at around 15 meV, indicated by the translucent line in Fig.~\ref{fig:phonons}(a). This excitation is most pronounced at the lowest temperature, and is observed in both RR and  RL polarizations (see Figure \ref{fig:magnon}(a)). As we warm up above  $T_{\rm N} \approx 50$~K, this excitation shifts towards lower energy, and shows a broad maximum around 12.4 meV(100 cm$^{-1}$) and finally disappears by 121~K. The temperature dependence of the frequency and the linewidth of this excitation is summarized in Fig. \ref{fig:magnon_summary} of \hyperref[sec:SI]{SI}.  The decreasing frequency and broadening of the excitation with increasing temperature are consistent with the decrease of the coherence length of the order above $T_{\rm N}$.  Below, we show that it can be assigned to 2M excitations of ordered V spins. 

Magnetic Raman scattering in antiferromagnets typically produces a 2M excitation, which can be understood within an FL process \cite{Fleury1968}. This process describes an excitation exchanging electrons between neighboring  sites of the two sublattices of an antiferromagnet with the Raman tensor
 $R = \sum_{\langle ij \rangle} (\vec{e}_\mathrm{in} \cdot \hat{\delta}_{ij}) (\vec{e}_\mathrm{out} \cdot \hat{\delta}_{ij}) \vec{S}_i \cdot \vec{S}_j$. 
 Here $\vec{e}_\mathrm{in}$ and $\vec{e}_\mathrm{out}$ are polarizations of excitation and scattered light, and $\vec{\delta}_{ij}$ is a vector connecting $\vec{S}_i$ and $\vec{S}_j$ sites of the two AFM sublattices. In \vns, non-symmorphic unit cell and positions of V atoms ensure a non-zero projection of $\vec{\delta}_{ij}$ on the $(ab)$ plane resulting in the fact that we observe FL process in the geometry, where  $\vec{e}_\mathrm{in}$ and $\vec{e}_\mathrm{out}$ lie in the $(ab)$ plane. As Fig. \ref{fig:magnon}(a) shows, we observe 2M excitation in both  RR and RL scattering channels,  reflecting the hexagonal AFM structure observed in the in-plane projection (Fig. \ref{fig:magnon}(b)).  Due to the very low intensity of the spectra in RL polarization, it was possible to obtain the spectra only with 15 mW laser power, which resulted in 19 K temperature of the sample. The comparison with RR shows the presence of the 2M feature in both channels, with some intensity in RL at lower frequencies due to the elevated temperature, obscuring the linear dependence of the electronic Raman scattering continuum on frequency.

While FL process is allowed to be short range \cite{Fleury1968}, in an ordered state the 2M energy scale can be understood by comparing it to a linear spin wave theory, where the magnon dispersion obtained from the calculations is integrated over the entire Brillouin zone  (BZ) selecting 2M states with $\vec{k}_1 + \vec{k}_2 = 0$  \cite{Tennant2003}. We perform spin wave calculations for \vns,  assuming an AFM $A$-type structure with spins pointing along the crystallographic $a$-axis, with FM order in the layers and AFM order between the layers, depicted in Fig. \ref{fig:struc}(a) \cite{Youzhe2025}. We use a minimal model of FM nearest-neighbor exchange $J_1<0$ and next-nearest-neighbor AFM exchange $J_2>0$, as shown in Fig.  \ref{fig:struc}(b). In this structure, each site has six in-plane nearest neighbors with FM interactions and six out-of-plane nearest neighbors connected by AFM exchange interactions. We perform linear spin wave theory calculations using a model Hamiltonian 
 \begin{equation}
H = \sum_{ij} J_{ij} \vec{S}_{i} \cdot \vec{S}_{j} + \sum_{i} D_z (S_{i}^z)^2
 \end{equation}
where $i,j$ are the indices for the two sites of the unit cell, and $D_z$ is the single-ion anisotropy that disfavors spins lying out of the basal plane. We use a $J_1-J_2$ model, assuming V$^{3+}$ has quenched angular momentum and spin quantum number $S=1$, and estimate the exchange parameters $J_1 = -0.55$ meV and $J_2 = 0.75$ meV from the inelastic neutron scattering spectrum, which will be discussed elsewhere \cite{Chris2025}. The single ion anisotropy $D_z$ is estimated to be less than 0.1 meV. 

The details of the calculations of the 2M scattering, which allow for comparison to Raman scattering results, are discussed in \hyperref[sec:SI]{SI}. The calculation with the parameters listed above produces the 2M spectrum in the energy range, which is overall in good agreement with the Raman scattering experimental spectra, as demonstrated in Fig.~\ref{fig:magnon}(a). The peak features are correlated with extremal magnon energies at different points of the BZ boundary. In particular, the lower-energy $\sim 7$ meV peak occurs due to the van Hove singularity of the {correlation with A-point} out-of-plane magnons with energy $6SJ_2$, whereas the largest $\sim 18$ meV peak occurs due to in-plane K-point magnons with energy $S(6J_2 - 9 J_1)$. Our model is able to demonstrate qualitatively the presence of 2M excitations in two distinct polarization channels with slightly varying intensity (see Fig.~\ref{fig:magnon_new} (a)), consistent with our experiment. Differences in the peak positions and the distribution of the spectral weight between RR and RL polarizations arise from differing weights in momentum space probed by the Raman tensor.  Intuitively this result can be understood within FL model, where incident $\vec{e}_\mathrm{in}$ and scattered $\vec{e}_\mathrm{out}$ light electrical field vectors should be parallel to $\vec{\delta}_{ij}$, the vector connecting two AFM sublattices: in the geometry of the ordering suggested in Fig. \ref{fig:struc}(b), a projection of the collinear ordered AFM on the $(ab)$ plane is a hexagon of V atoms connected by AFM interactions, with different polarizations of the in-$(ab)$-plane electric field selecting different components of this projection.

While the experimentally measured high-frequency cut-off of the magnetic band is well reproduced by the calculations, the maximum related to van Hove-like singularity appears broader and shifted to a lower frequency in the experimental spectra. One possible explanation for this discrepancy includes a deviation from $\vec{k}_1 + \vec{k}_2 = 0$ momentum conservation rule. Such a deviation could arise from breaking translational symmetry, such as by a stacking fault.  It would affect the interaction between adjacent planes and therefore strongly couple to $J_2$, which controls both the bandwidth of the out-of-plane and in-plane magnon energies and is the origin of both the weak peak around 7 meV and the sharp peak around 18 meV in the calculated 2M spectra. The FL process is the most sensitive to $J_2$. The variation in the energy scale $J_2$ could then shift the spectral weight and cause the broadening and smoothing of the 2M peak on both ends, as observed in the data.

\par To conclude, our Raman data on the presence of 2M scattering confirm the antiferromagnetic component of the magnetic order, rejecting possibilities of purely ferromagnetic order~\cite{Parkin1980,Inoshita2019, Hawkhead2023}. The energy and polarization dependence of the observed 2M scattering are in agreement with the A-type order and neutron scattering results~\cite{Youzhe2025}  and the most recently reported calculations, based on a systematic first-principles study \cite{Hatanaka2023}.

{\it Electronic Raman scattering as the probe of electronic structure} Electronic Raman scattering continuum is identified in room temperature measurements by comparing in- and out-of-$(ab)$ plane scattering channels (Fig. \ref{fig:outofplane}). The continuum scattering $\chi''_{el}(\omega) \propto \omega$ is present only in the $(xx)+(xy)$ channel, indicating a highly 2D character of the electronic structure of the 3D layered V$_{1/3}$NbS$_2$, in agreement with high electronic anisotropy from resistivity measurements (see SI in~\cite{Youzhe2025}). A similar conclusion is suggested by the presence of electron-phonon coupling for $A_1$ modes, seen only in the in-plane scattering.  The slope of the $\chi''_{el}(\omega) \propto \omega$ extends up to $\approx$ 62~meV. It is followed in the RR spectra down to 14~K and doesn't show any overall change upon cooling from 121 K to 14 K. Raman scattering excited with both 514.5 nm and 488 nm laser lines (Fig.~\ref{fig:Extrange}) not only confirms the origin of the continuum, but also demonstrates that above approximately 150 meV the frequency dependence changes. 

The linear in frequency continuum of electronic excitations can be understood by comparing it with the DMFT calculations of the electronic band structure in the paramagnetic state of \vns~\cite{Youzhe2025}, with similar results suggested by calculations in a FM state \cite{Inoshita2019},  which also reinforce the conclusions of these calculations.  The calculations suggest that \vns\ is a semimetal with tilted Weyl nodes close to $E_{\rm F}$  found between $\Gamma$ and K and $\Gamma$ and M points of the BZ, protected by the absence of inversion symmetry in the paramagnetic state. Raman scattering can excite vertical interband transitions in semimetals (see the inset in Fig. \ref{fig:phonons}(a)), producing a continuum of interband excitations \cite{Niko2024}. The frequency dependence of this electronic continuum is determined by the shape of the dispersion and dimensionality of the electronic structure: It was shown in  Ref.~\cite{Niko2024} that the general form for the frequency dependence of the continuum of Raman electron-hole interband excitations is $\chi''(\omega) \propto \omega^{2(l-1)+d-3}$, where $l$ = 2 is power characteristic of a Dirac or Weyl node, and $d$ is dimensionality. For example, such linear Raman scattering continuum of interband electron-hole excitations was reported in a gated single-layer graphene device~\cite{Riccardi2016}, while in Nd$_2$Ir$_2$O$_7$ a Weyl node in 3D electronic structure resulted in $\chi''_{el}(\omega) \propto \omega^2$. Therefore, a linear dependence on frequency $\chi''(\omega) \propto \omega$ of the electronic continuum observed in \vns\ provides evidence for the predicted Weyl node, not obscured by other electronic bands and 2D character of the electronic structure. The linear continuum extends down to 1.7~meV, putting the high limit of the chemical potential at $\mu$=0.85~meV. 

In \vns, while Weyl nodes are protected by the absence of inversion symmetry above $T_{\rm N}$, time-reversal symmetry is additionally broken at temperatures below $T_{\rm N}$.  However, we do not observe an overall change in the slope in the linear dependence of the electronic scattering within the precision of our measurements.  

The linear dependence of the continuum with a singular slope value extends up to $\approx$ 62 meV (500 $cm^{-1}$)(Fig. \ref{fig:phonons}),  suggesting an upper-frequency cutoff for the spectral range where only transitions within Weyl nodes are observed. An even more notable change of the slope and increase of $\chi''(\omega)$ Raman intensity is observed above 150~meV (Fig. \ref{fig:Extrange}) and would be associated with electronic transitions between the bands with a higher density of states. While direct gaps between quadratically dispersed bands in the electronic band structure calculations ~\cite{Youzhe2025,Inoshita2019} are of the order of 0.5 eV, there is a prediction of flat bands close to  $E_{\rm F}$. This energy range makes them good candidates for the observed Raman intensity, assuming electronic transitions involve these flat bands.  Since the absence of inversion symmetry relieves a parity-selective selection rule for interband transitions, the electronic Raman scattering spectra will be integrating over all possible vertical electronic transitions, and the frequency dependence of the $\chi''(\omega)$ electronic continuum will be the result of the sum of all possible excitations at a given energy.

In conclusion, our Raman scattering study of \vns\ reveals electronic scattering consistent with the 2D electronic structure and the presence of Weyl nodes close to the $E_{\rm F}$, where the linear electronic bands dispersion extends up to 62 meV. We detect a magnetic ordering transition at 50~K, with observed 2M excitations consistent with $A-$type structure, where V spins are ordered ferromagnetically in $(ab)$ planes, and the planes are ordered antiferromagnetically. 
Our findings are in agreement with the Weyl semimetal state in \vns\ and indicate that the large Berry curvature associated with the Weyl points near the $E_{\rm F}$ generates large transverse responses observed in the collinear AFM state of this material. It is the first experimental confirmation of a collinear AFM Weyl semimetal and opens up avenues for enhancing the anomalous transverse responses in collinear antiferromagnets, including the altermagnets.
Our work serves to highlight Raman scattering as a powerful tool for not only studying magnetic excitations but also exploring the consequences of nontrivial topology in electronic band structures.

\begin{acknowledgments}
The authors are thankful to C. Broholm, P. Nikolic, and  Y. Yang for useful discussions. This work was supported as part of the Institute for Quantum Matter, an Energy Frontier Research Center funded by the U.S. Department of Energy, Office of Science, Basic Energy Sciences under Award No. DE-SC0019331. Work in Japan was supported by JST-ASPIRE(JPMJAP2317) and JST-Mirai (JPMJMI20A1) Programs.
\end{acknowledgments}

\bibliography{VNS_12122024}

\clearpage
\onecolumngrid

\begin{center}
{\LARGE Supplementary Material: Raman spectroscopic evidence for linearly dispersed nodes and magnetic ordering in the topological semimetal  V$_{1/3}$NbS$_2$} \\[1ex] 
{\large Shreenanda Ghosh$^{1}$, Chris Lygouras$^{1}$, Zili Feng$^{2,3,4}$, Mingxuan Fu$^{2,3}$, Satoru Nakatsuji$^{1,2,3,4}$, Natalia Drichko$^{1}$}

\noindent{\textit{$^{1}$Institute for Quantum Matter and William H. Miller III Department of Physics and Astronomy, Johns Hopkins University, Baltimore, Maryland 21218, USA}}\\

\noindent{\textit{$^{2}$ Department of Physics, University of Tokyo, Bunkyo-ku, Tokyo 113-0033, Japan}}\\

\noindent{\textit{$^{3}$Institute for Solid State Physics, University of Tokyo, Kashiwa, Chiba 277-8581, Japan}}\\

\noindent{\textit{$^{4}$ Trans-scale Quantum Science Institute, University of Tokyo, Bunkyo-ku, Tokyo 113-0033, Japan}}\\

\end{center}

\section{} \label{sec:SI}

\twocolumngrid
\subsection{Synthesis}
The measured V$_{1/3}$NbS$_2$ single crystals were synthesized using chemical vapor transport with iodine as a transport agent. Polycrystalline samples were first made by heating high-purity vanadium, niobium, and sulfur at 1000$^\circ$C for 24 hours. After cooling, 2 g of the powder was mixed with 0.1 g of iodine in an evacuated silica tube and heated for 7 days in a three-zone furnace.  The crystals are shaped like hexagonal plates with the most developed face of about 1 mm by 1 mm parallel to the $(ab)$ crystallographic plane, with a thickness of about 50 $\mu$m.

\subsection{Raman scattering experiments}

Polarization-resolved Raman scattering measurements in the temperature range 14-300 K were performed using single crystals of V$_{1/3}$NbS$_2$.  
Raman scattering measurements were done using a T64000 Horiba-Jobin-Yvon spectrometer with a LN$_2$ cooled CCD in two different configurations. To perform Raman measurements probing the out-of-$(ab)$-plane and to compare it with in-$(ab)$-  spectra  
at 300 K in the 133-700 cm$^{-1}$ range, a micro-Raman option of the spectrometer equipped with an Olympus microscope was used. The laser probe was 2 $\upmu$m in diameter. The 514.5 nm line of Ar+-Kr+ mixed gas laser was used as the excitation light with the laser power not exceeding 500 $\upmu$W, to avoid heating and laser damage to the sample. The spectral resolution was 2 cm$^{-1}$. The out-of-$(ab)$-plane and  in-$(ab)$- plane spectra were collected from two different samples grown in an identical method. 

The Raman spectra in the $(ab)$ plane at temperatures between 300 and 14 K in the 13-670 cm$^{-1}$ spectral range were collected using the macro configuration of the spectrometer in the pseudo-Brewster’s angle geometry with an elliptical laser probe of 50 $\times$ 100 $\upmu$m in size. Laser lines 514.5 nm and 488 nm were used for excitation,  with the laser power varied between 10 - 15 mW, and the estimated heating of the sample of 1 K per 1 mW of laser power. The reported temperature values include the estimated laser heating.
 For measurements below 300 K, the sample was mounted on the cold finger of a Janis ST-500 cryostat using GE varnish.

The spectra in Fig. \ref{fig:outofplane} and Fig. \ref{fig:phonons} have been normalized to the laser power used for the measurements.  Dark current was subtracted from each spectrum, and the spectra presented in Fig. \ref{fig:Extrange}, were normalized by the spectrometer response. The presented Raman response $\chi''(\omega, T)$ was normalized on the Bose-Einstein factor $[\eta(\omega,T)+1]$, where $\eta(\omega,T) = [exp (\frac{\hbar\omega}{K_BT}) -1 ] ^{-1}$ is the Bose occupation factor.

\subsection{Polarizations and symmetry analysis}

V$_{1/3}$NbS$_2$ belongs to the hexagonal space group P6$_3$22  which corresponds to $D_6$ point group \cite{Parkin1980, Hall2021}. The relationship between the measurement geometry and scattering channels for the $D_{6}$  point group of V$_{1/3}$NbS$_2$ is listed in Table \ref{table:geometries}.

\begin{table}[!htbp] \centering
\begin{tabular}{@{\extracolsep{5pt}} lll}
\\[-1.8ex]\hline
\hline \\[-1.8ex]
\textbf{Scattering Geometry} & \textbf{irreps}  \\
\hline \\[-1.8ex]
\hline \\[-1.8ex]
XX &  $A_1 + E_2$ \\

\hline

ZZ & $A_1$ \\

\hline

RR ($ab$ plane) & $A_1$ \\

\hline

RL ($ab$ plane) & $E_2$ \\

\normalsize
\end{tabular}
\caption{Relationship between the scattering geometries and the corresponding irreducible representations.}\label{table:geometries}
\end{table}
\vspace*{-\baselineskip}

To compare in- and out-of $(ab)$ plane spectra, measurements were performed using the micro-Raman setup in backscattering geometry.  The in-$(ab)$-plane of the crystals, the spectra were collected in $(xx)+(xy)$ channels,  where $x$ direction was parallel the (1$\bar{1}$0) crystallographic direction. For out-of-plane measurements, we used a naturally grown face of the crystal 50-200 $\upmu$m wide, found at an angle of about 70 degrees to the $(ab)$ plane.  Therefore, unlike the schematic diagram shown in the inset of Fig. \ref{fig:outofplane}, the probed face of the as-grown crystals in out-of-$(ab)$-plane measurements, is not exactly perpendicularly aligned to the $(ab)$ plane, and the resulting spectra probe $(zz)+(zx)$ scattering channel.

The spectra in macro pseudo-Brewster's angle configuration were measured in circular polarizations from the $(ab)$ plane as discussed in the main text.

\subsection{Raman data analysis}

Full symmetry analysis for Raman active phonons in \vns\ spectra is presented in Table~\ref{table:Character table}. Observed phonons discussed in the main text were assigned according to this symmetry analysis. 

\begin{table}[!htbp] \centering
\begin{tabular}{@{\extracolsep{5pt}} lll}
\\[-1.8ex]\hline
\hline \\[-1.8ex]
\textbf{Wyckoff positions} & \textbf{irreps}   & \textbf{ No. of Raman} \\

\newline
\textbf{(Atom)} & \textbf{(Functions)} & \textbf{active phonons}\\
\hline \\[-1.8ex]

\hline \\[-1.8ex]
12i(S) &  $A_1$(x$^2$ + y$^2$, z$^2$) & \quad 3\\

& $E_2$(x$^2$ - y$^2$, xy) & \quad 6\\

& $E_1$(xz,yz) & \quad 6\\

2a(Nb2) &  $E_2$  & \quad 1 \\

& $E_1$   & \quad 1 \\

\hline

4f(Nb1) &  $A_1$ & \quad 1\\

&  $E_2$  & \quad 2 \\

& $E_1$   & \quad 2 \\

\hline

2c(V) &  $E_2$  & \quad 1 \\

& $E_1$   & \quad 1 \\

\normalsize
\end{tabular}
\caption{Wyckoff positions for each atom in V$_{1/3}$NbS$_2$ and the corresponding irreps $A_1$, $E_1$, and $E_2$ of the D$_{6}$ point group. The functions that are Cartesian products transforming as the various irreps of the group are shown within the parentheses. The third column shows the number of expected Raman active phonons.}\label{table:Character table}
\end{table}

\begin{figure}
\includegraphics[width=\columnwidth]{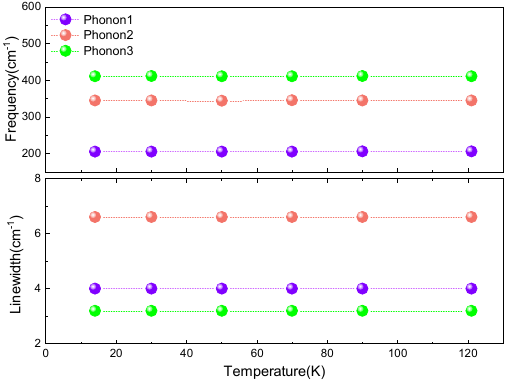}
\caption{Temperature dependence of the peak frequencies (top) and linewidths (bottom) of the $A_1$ phonons.}\label{fig:parameters}
\end{figure}

To obtain temperature dependence, we fit the spectra presented in Fig. \ref{fig:phonons} with a combination of different line shapes and a linear background. The line shape of the $A_1$ phonons at 202 cm$^{-1}$ and 406 cm$^{-1}$ were well described by a characteristic asymmetric Fano line shape whereas the $A_1$ phonon at 375 cm$^{-1}$ is well described by a Voigt peak. Fano shape is a typical result of an interaction between a single level of a phonon mode and an underlying continuum of excitations and can be described by the formula

\begin{equation}
F(\omega, \omega_F, \Gamma_F, q) = \frac{1}{\Gamma_F q^2} . \frac{[q+ \alpha (\omega)]^2}{[1 + \alpha(\omega)^2]}
\end{equation} where $\alpha(\omega)= \frac{\omega -\omega_F}{\Gamma_F}$ and q defines the coupling to the continuum \cite{Fano1961, Zhang2015}.

\begin{figure}
\includegraphics[width=\columnwidth]{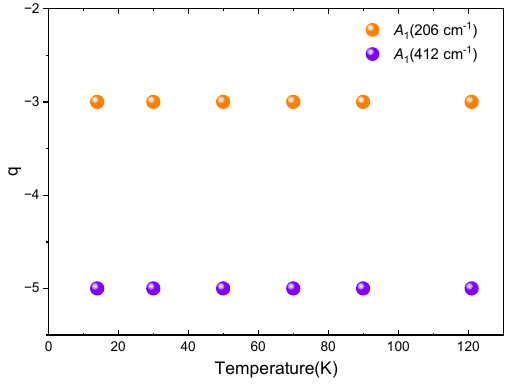}
\caption{Temperature dependence of the Fano parameter for two out of the three $A_1$ phonons.} \label{fig:q}
\end{figure}

Since for the in-plane measurements, the phonons show a Fano shape with a small Fano factor (Fig. \ref{fig:q}), therefore it was possible to fit them also with a Voigt function to compare to the out-of-plane phonons: Each of the phonons in Fig.~\ref{fig:outofplane} were fitted using a Voigt function, to reliably compare the widths. The spectrum corresponding to out-of-plane phonons has been normalized by the in-plane intensity of the $A_1$ phonon at 406 cm$^{-1}$, for an accurate comparison of the background contribution.

As discussed in the main text, $A_1$ phonons show very weak temperature dependence. To illustrate that, we plot the temperature dependence of the position and line width of the phonons in Fig.~\ref{fig:parameters}. Fano coupling is temperature-independent as well, which is demonstrated in Fig~\ref{fig:q}.

$A_1$ phonons at 202 cm$^{-1}$ and 406 cm$^{-1}$ show a broader polarization-independent wing (Fig. \ref{fig:magnon_new}). Typically, such features are a signature of stacking faults \cite{Rohmfeld1999}, which is a natural possibility for a layered crystal with weak bonds between the layers. 

As discussed in the main text, in $(zz)$ polarization, we observe $E_1$ symmetry phonons. Frequencies and line widths of these phonons are presented in Table~\ref{table:E1phonons}.

\begin{table}[!htbp] 
\centering
\small 
\setlength{\tabcolsep}{5pt} 
\renewcommand{\arraystretch}{1.3} 
\begin{tabular}{@{\extracolsep{4pt}} ccc}
\\[-1.8ex]\hline
\hline \\[-1.8ex]
\thead{Frequency \\ (cm$^{-1}$)} & \thead{Symmetry} & \thead{linewidth in ZZ \\ (cm$^{-1}$)} \\ 
\hline \\[-1.8ex]
273 & $E_1$ & 3 \\ 
304 & $E_1$ & 4 \\ 
323 & $E_1$ & 5 \\ 
356 & $E_1$ & 5 \\ 
375 & $E_1$ & 5 \\ 
388 & $E_1$ & 4 \\ 
\hline
\hline
\end{tabular}
\caption{Frequencies and linewidths for the $E_1$ phonons observed in the out-of-plane configuration.}
\label{table:E1phonons}
\end{table}


\begin{figure}
\includegraphics[width=\columnwidth]{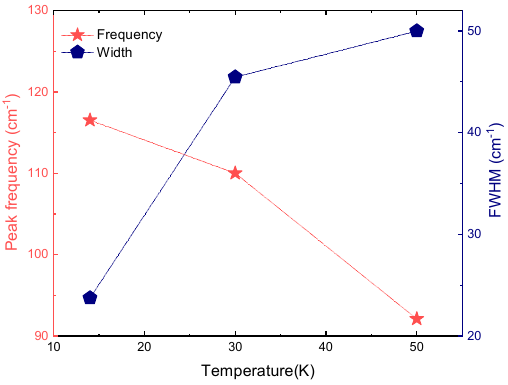}
\caption{Temperature dependence of the peak frequency and linewidth of the 2M excitation.} \label{fig:magnon_summary}
\end{figure}

\subsection{Spin wave calculations}
Magnetic susceptibility measurements indicate near-perfect cancellation of the moments between the magnetic sublattices of the AFM structure in V$_{1/3}$NbS$_2$, which is not significantly modified by the application of a magnetic field \cite{Hall2021}. This almost perfectly collinear AFM structure (except a small canting out of the basal plane, resulting in a small net moment observed in magnetization) will certainly produce 2M excitations as a result of flipping of a pair of spins. Note that a $\Delta S_x = 0$ process would require spin flip processes in adjacent layers that are antiferromagnetically aligned, rather than within a single FM layer. 

The 2M calculation requires details of the spin wave modes. These can be estimated from linear spin wave theory. We consider the Fourier transform of the exchange interactions, where $J_A(\vb{q})$ are for interactions between atoms in the same plane, and $J_{AB}(\vb{q})$ are the interactions between adjacent planes. Following the conventional Holstein-Primakov approximation of the spin operators, the isotropic Heisenberg Hamiltonian with single-ion anisotropy can be expressed as $H= \sum_{\vb{q}} \frac{1}{2} \vb{X}_{\vb{q}}^\dagger \mathsf{H}_{\vb{q}} \vb{X}_{\vb{q}}$ where $\vb{X}_{\vb{q}} = (a_{\vb{q}}, b_{\vb{q}}, a^\dagger_{-\vb{q}}, b^\dagger_{-\vb{q}})^T$ is the vector of the Bose operators for the two sites. In our $J_1-J_2$ model on the triangular lattice, we have that 
 \begin{align}
     J_A(q) &= 2J_1 [\cos 2\pi h + \cos 2\pi k + \cos 2\pi(h+k)]  \\ 
  |J_{AB}(q)|^2 &= 4J_2^2 \cos^2\pi l [ 8\cos^2\pi h + 8 \cos^2 \pi k \nonumber \\
    &\quad - 8 \sin \pi h \sin \pi k \cos\pi(h+k) - 7 ] 
 \end{align} 
and by diagonalizing the Hamiltonian \cite{Boothroyd2020, Huberman2005}, the energies of the one-magnon states are 
 \begin{equation}
     \omega_\pm(\vb{q}) = S\sqrt{ [J_A(\vb{q}) + D_z - J_A(0) + J_{AB}(0) ]^2 - [D_z \pm |J_{AB}(\vb{q})|]^2 } 
 \end{equation}
The values of the exchange interactions for arbitrary-distance interaction scale can be estimated from the inelastic neutron scattering spectrum, which will be the work of a forthcoming publication. 

Within the framework of the FL theory \cite{Fleury1968, Perkins2008, Yang2022}, we calculate the polarization dependence of the 2M scattering intensity using the Raman tensor 
 \begin{equation}
     R = \sum_{\langle ij \rangle} (\vec{e}_\mathrm{in} \cdot \hat{\delta}_{ij}) (\vec{e}_\mathrm{out} \cdot \hat{\delta}_{ij}) \vec{S}_i \cdot \vec{S}_j 
 \end{equation}
where we assume that the interaction between spins is isotropic (Heisenberg-like), and $\langle ij \rangle$ indicates the sum over the next-near-neighbor spins (coupled by $J_2$) that are the first pair aligned antiferromagnetically. Taking a Fourier transform, we express the spin operators using ladder operators $A_{\vec{k}}$ within linear spin wave theory using the Holstein-Primakoff transformation, and diagonalize these using a Bogoliubov transformation $A_{\vec{k}} = S_{\vec{k}} \widetilde{A}_{\vec{k}}$. For 2M scattering, we consider the tensor $\widetilde{R}({\vec{k}}) = S^\dagger_{\vec{k}} \widetilde{R}_0(\vec{k}) S_{\vec{k}}$ where $\widetilde{R}_0(\vec{k})$ is a 4x4 tensor related to the Fourier transform $\sum_j (\vec{E}_\mathrm{in} \cdot \hat{\delta}_{ij}) (\vec{E}_\mathrm{out} \cdot \hat{\delta}_{ij}) e^{i\vec{k}\cdot \vec{\delta}_{ij}}$. The scattering intensity is proportional to the imaginary part of the Green's function, 
 \begin{equation} 
     I(\omega) \propto -\Im \sum_{\nu,\mu=1,2} \sum_{\vec{k}} \frac{ \widetilde{R}_{\mu+2, \nu}(\vec{k}) \widetilde{R}_{\mu,\nu+2}(\vec{k}) }{\omega-(\omega_\mu + \omega_\nu)+i\epsilon} 
 \end{equation} 
where the sum $\nu,\mu$ is over the 2M branches, and the sum over wave vector is taken over the first BZ. The polarization dependence is deduced by using normalized electric field vectors expressed in the lab frame. 

Fig. \ref{fig:magnon_summary} summarizes the temperature dependence of the frequency
and the linewidth of the two-magnon excitation. These parameters
were obtained by fitting the two-magnon peak using
a Gaussian profile.

\end{document}